\documentclass[pra,twocolumn,amssymb,amsmath,showpacs,aps,floatfix]{revtex4}
\newcommand{\be}{\begin{equation}}
\newcommand{\ee}{\end{equation}}
\newcommand{\bea}{\begin{eqnarray}}
\newcommand{\eea}{\end{eqnarray}}

\usepackage{graphicx}
\usepackage{epsfig}
\usepackage{bbold}

\begin{document}

\title{Artificial Gauge Field for Photons in Coupled Cavity Arrays}
\author{R. O. Umucal\i lar}
\email{onur@science.unitn.it}
\affiliation{INO-CNR BEC Center and Dipartimento di Fisica, Universit\`a di Trento, I-38123 Povo, Italy}
\author{I. Carusotto}
\email{carusott@science.unitn.it}
\affiliation{INO-CNR BEC Center and Dipartimento di Fisica, Universit\`a di Trento, I-38123 Povo, Italy}

\date{\today}

\begin{abstract}
We propose and characterize solid-state photonic structures where light experiences an artificial gauge field.
A non-trivial phase for photons tunneling between adjacent sites of a coupled cavity
array can be obtained by inserting optically active materials in the structure or by inducing a suitable coupling of the propagation and polarization degrees of freedom.
We also discuss the feasibility of observing strong gauge field effects in the optical spectra of realistic systems, including the Hofstadter butterfly spectrum.
\end{abstract}


\maketitle

\section{Introduction}

The effect of an external magnetic field on the dynamics of charged particles underlies a number of intriguing phenomena in very different contexts, ranging from magnetohydrodynamics in astro- and geo-physics to the fractional quantum Hall effect in solid state physics~\cite{FQH_review}.
Upon quantization, the eigenstates for non-interacting particles in a uniform magnetic field form a simple equispaced ladder of highly degenerate Landau levels in free space, while the interplay with a periodic lattice potential was predicted to give rise to fractal structures in the energy vs. magnetic flux plane, the so-called Hofstadter butterfly~\cite{Hofstadter}. So far, experimental observation of such a fascinating structure in ordinary solids has been hindered by the extremely high value of the required magnetic field intensity~\cite{Albrecht}.

In recent years, an intense theoretical activity has investigated the possibility of generating artificial gauge fields for neutral atoms by taking advantage of the Berry phase~\cite{Berry} accumulated by an optically dressed atom which adiabatically performs a closed loop in real space~\cite{olshanii,Dalibard}: the nucleation of a few quantized vortices in a Bose-Einstein condensate under the effect of an artificial gauge field has been demonstrated in the pioneering experiment by Lin {\em et al.}~\cite{spielman}.
The combination of a gauge field with atom-atom interactions is expected to give rise to strongly correlated atomic gases that closely remind quantum Hall liquids~\cite{FQH}.

In the meanwhile, experimental advances in the generation and manipulation of photon gases in semiconductor devices have opened the way to the study of collective many-body effects in quantum fluids of light~\cite{EPN}. The first reports of Bose-Einstein condensation~\cite{PhotBEC} have been recently followed by the demonstration of superfluid flow around defects~\cite{Pol_Superfl_Exp} and the hydrodynamic nucleation of vortices and solitons~\cite{Pol_Vort_Exp}.

In this work, we theoretically investigate photonic devices where the orbital motion of the photon experiences an artificial gauge field. Previous work in this direction has considered arrays of coupled optical cavities confining single atoms~\cite{atoms}, topological electromagnetic states in gyromagnetic photonic crystals~\cite{PC}, and time-reversal symmetry breaking effects for microwaves in circuit-QED devices~\cite{Girvin}.
In contrast to these works, our scheme can be implemented with standard solid-state photonic technology in the visible or infrared spectral range~\cite{Hafezi}. In combination with the on-going research on strongly correlated photon systems~\cite{TG}, it is expected to open new perspectives in the study of non-equilibrium many-body physics under strong magnetic fields.

The basic idea of our proposal consists of imposing a non-trivial tunneling phase to photons by taking advantage of the polarization degree of freedom: this phase can be generated by an optically active medium embedded in the structure or can have a geometric nature. Geometric phases have been demonstrated in a number of configurations for propagating light~\cite{Berry_optics,pancha}. Here, we extend the idea to the case where photons are confined in a two-dimensional lattice and the phase is acquired by an evanescent wave while tunneling between neighboring sites. Two classes of devices, which can be built using passive dielectric materials with a real refractive index, are specifically considered.

The first configuration is illustrated in Sec.~\ref{sec:first} and is based on an array of optical cavities: this scheme is suitable for observing general gauge field effects on photons in the non-interacting regime and shows the interesting possibility of scaling the structure to any wavelength region.
The second configuration is presented in Sec.~\ref{sec:second} and is based on a single planar microcavity with a periodic lateral patterning. This scheme appears as most promising in view of combining the artificial gauge field with strong optical nonlinearities, so as to enter the regime of strongly correlated photon gases.
To complete the study, Sec.~\ref{sec:observables} is devoted to the discussion of some observable quantities that can be used to extract the physics of quantum particles in strong gauge fields from experimentally accessible optical spectra.

\section{The first scheme: array of DBR microcavities}
\label{sec:first}

\begin{figure}[htbp]
\includegraphics[scale=0.55]{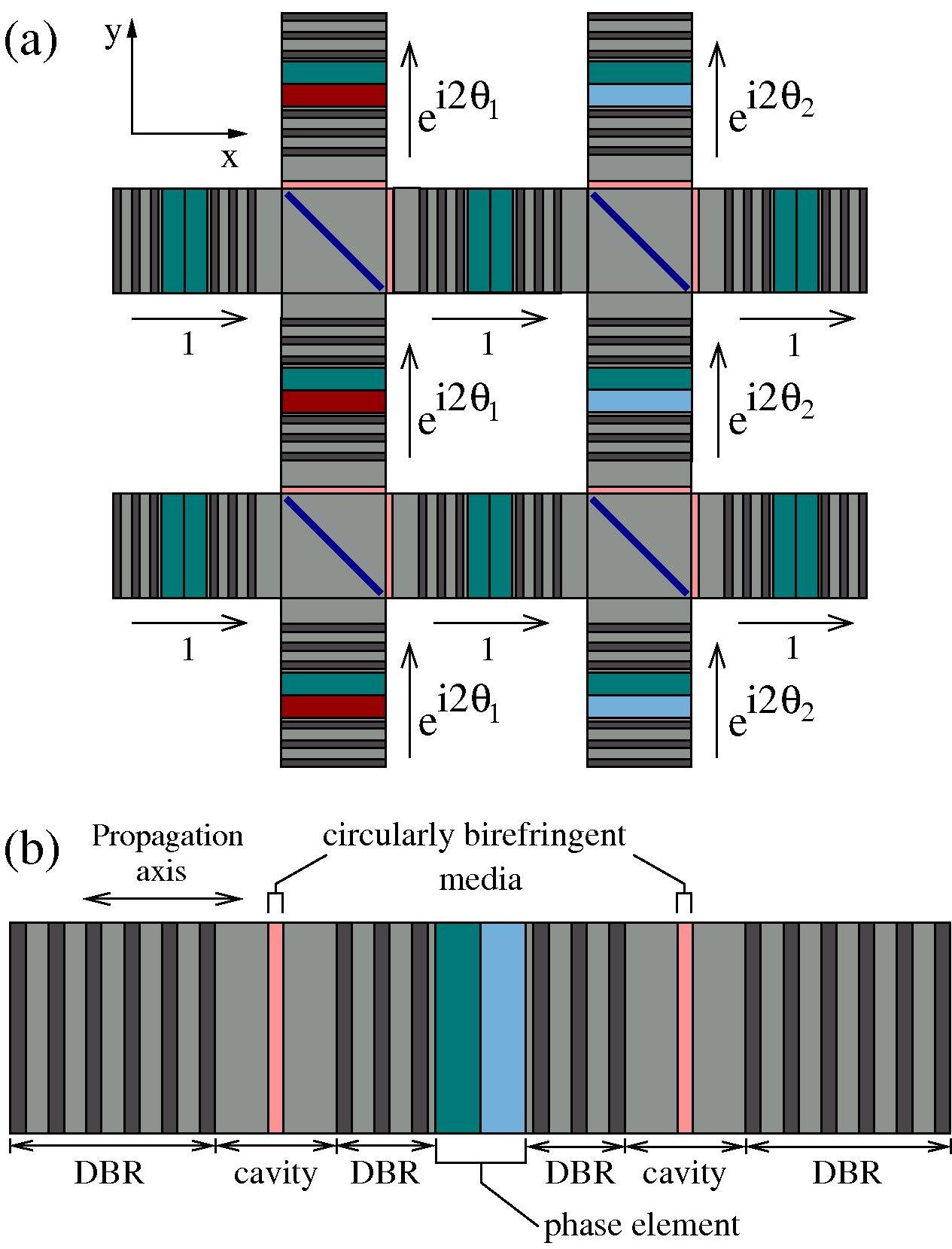}
\includegraphics[scale=0.4]{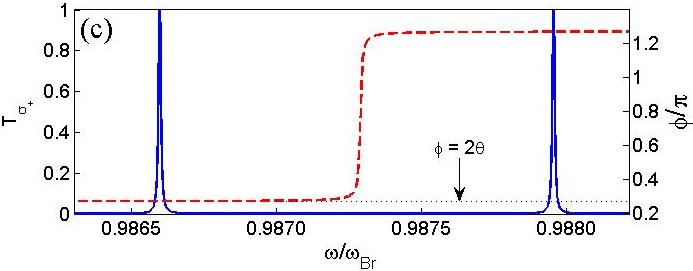}
\caption{(Color online) Scheme of the coupled DBR microcavity configuration to generate the artificial magnetic field. (a) Two-dimensional square lattice of cavities: in the sketch, four cavities containing an oblique mirror each are connected by phase elements embedded between DBR mirrors. Within the Landau gauge, the artificial magnetic field arises from the $x$-dependence of the tunneling phase in the $y$ direction. (b) Simplest two-cavity set-up. (c) For the two-cavity set-up of panel (b), transmission spectrum (blue solid line) for $\sigma_+$ incident light; relative phase $\phi$ of the field in the two cavity layers (red dashed line).
The outer (inner) DBRs contain 19 (10) periods of alternate layers of refractive index $n_{1}=3.6$ and $n_2=2.9$ and optical thickness $\lambda_{\rm Br}/4$, where $\lambda_{\rm Br}=2\pi c/\omega_{\rm Br}$, $\omega_{\rm Br}$ being the Bragg frequency. The cavity layers are $\lambda_{\rm Br}/2$ thick and have $n_{\rm cav}=2.9$. They also contain a circularly birefringent medium of thickness $d_{\sigma}/\lambda_{\rm Br} = 0.21/\pi$ with $n_{\sigma_\pm} = 1.2, 2$. The angle $\theta$ between the optical axes of the half-wave slabs is arbitrarily chosen to be $3\pi/22$. The same effect is observed in the presence of an optically active slab embedded in between the inner DBRs.
\label{fig:Schematic_setup}}
\end{figure}

This first configuration is sketched in Fig. \ref{fig:Schematic_setup}(a) and consists of a two-dimensional array of optical cavities. Even if a similar physics can be observed in a wide class of systems, for concreteness we shall focus our attention on the specific case of monolithic distributed Bragg reflector (DBR) cavities~\cite{transm_mat}. Within each cavity, light propagates along four arms oriented along the $x,y$ axes, each of which is terminated by a DBR mirror. A polarization preserving, weakly reflecting mirror oriented at 45 degrees with respect to the $x,y$ axes is located at the center of each cavity and serves to mix light in the different arms: in this way, the photonic eigenmodes of each isolated cavity will be linear superpositions of two standing waves along the $x$ and $y$ axes. In the following we shall focus our attention on a single optical mode per cavity, e.g. the one consisting of a symmetric superposition of two standing waves along the two axes with a given circular polarization, say $\sigma_+$~\cite{footnote0}. In order to suppress unwanted mixing of the circularly polarized $\sigma_\pm$ states by the central, oblique mirror, suitable circularly birefringent layers
can be used to lift the degeneracy of $\sigma_\pm$ cavity modes~\cite{footnote1}. A more detailed discussion on the structure of the photonic modes within each cavity is given in the Appendix.

Coupling between neighboring cavities occurs via evanescent wave tunneling across the separating DBR mirrors. In order to generate the tunneling phase responsible for the artificial gauge field, two options can be envisaged.
The first one involves inserting a pair of linearly birefringent half-wave slabs within each DBR mirror separating neighboring cavities. The optical axes of the two slabs have a relative rotation angle $\theta$ around the propagation direction. Propagation through the first half-wave slab transforms the incident $\sigma_+$ light into $\sigma_-$ one; propagation through the second slab brings polarization back to $\sigma_+$, yet with an additional Pancharatnam phase factor $e^{2i\theta}$ of geometric nature~\cite{pancha}. On the other hand, $\sigma_+$ light tunneling through the mirror in the backward direction will acquire an opposite phase factor $e^{-2i\theta}$. Unwanted cavity-like resonances at the linearly birefringent slabs in the spectral vicinity of the cavity mode of interest can be ruled out by a careful choice of the slab parameters.

 The second choice involves a single slab of optically active medium in place of the pair of linearly birefringent slabs. The tunneling phase is in this case generated by the optically active material, which imposes phases $\pm\omega\,\Delta n\,d/2c$ to circularly polarized photons traveling across it in opposite directions.
 Here, $d$ is the thickness of the optically active slab and $\Delta n$ is the difference between the refractive indices experienced by the two helicity states.

To verify the existence of the tunneling phase when either a pair of linearly birefringent layers or an optically active layer is embedded within a DBR mirror, we performed transmission matrix calculations for the simplest two-cavity configuration schematically illustrated in Fig. \ref{fig:Schematic_setup}(b).
As one can see in Fig. \ref{fig:Schematic_setup}(c), transmission is maximum when the incident frequency is resonant with one of the two eigenmodes of the coupled cavity system. The presence of the non-trivial tunneling phase is apparent as a non-vanishing relative phase of the electric field in the two cavities: in standard configurations without any phase element, the two eigenmodes correspond to the symmetric and antisymmetric combinations of the isolated cavity modes, while now each of the eigenmodes exhibits an additional $\phi$ relative phase between the two cavities.

In a tight-binding model, the tunneling phase can be described in terms of a hopping Hamiltonian of the form $H_{\rm c}=-J e^{i\phi}\,\hat{b}_R^{\dagger}\,\hat{b}_L+\textrm{h.c.}$, with $J$ being a real and positive coefficient quantifying the strength of tunneling; $\hat{b}_{L,R}$ are the cavity mode operators for respectively the left and right cavities: it is straightforward to see that the eigenmodes of $H_c$ are indeed linear superpositions of the $\hat{b}_{L,R}$ isolated cavity modes with a $\phi$ phase difference between the two cavities.

The form of this coupling term directly extends to the full two-dimensional geometry shown in Fig. \ref{fig:Schematic_setup}(a), where the tunneling phase between each pair of neighboring cavities can be independently tuned by the relative orientation of the two half-wave slabs.
As usual, the hopping phase $\phi_{ij}$ between the neighboring $i,j$ cavities (equal to $\phi$ in the previous example) can be written in terms of an artificial gauge potential $\mathbf{A}$ as $\phi_{ij} = -\frac{e}{\hbar}\int^{\mathbf{r}_i}_{\mathbf{r}_j}\mathbf{A}\cdot d\mathbf{l}$, where $e$ is the elementary charge and the integral is performed along the segment connecting the cavities. A non-vanishing artificial magnetic field then appears in the photon dynamics whenever the sum of the tunneling phases around a closed loop is non-zero (modulo $2\pi$).

\section{The second scheme: laterally patterned planar DBR microcavity}
\label{sec:second}

The second scheme is based on a planar DBR microcavity architecture. Full three-dimensional confinement of the photon in micron-sized wells has been demonstrated by means of a lateral patterning of the cavity layer thickness~\cite{patterned_cav}. A periodic array of photon boxes can be obtained by means of a two-dimensional periodic repetition of the elementary well. As usual in Hubbard-like models, a tighter confinement within each well allows to enhance photon-photon interactions: with a sub-micron confinement, a strongly interacting photon regime is expected to be accessible using state-of-the-art semiconductor technology~\cite{TG,blockade}.

In addition, we assume a position-dependent vector field to be present that couples to the effective spin-1/2 system describing the photon polarization state in the $\sigma_{\pm}$ basis (defined with respect to the cavity growth axis $z$): the $z$-component of this field can result from a static magnetic field that splits the $\sigma_\pm$ polarization states by inducing a circular birefringence in the cavity material or by splitting the Zeeman components of an exciton state to which the photon is coupled~\cite{local_magnetic}.
On the other hand, the $x$ and $y$ components of the vector field (i.e. the ones that mix $\sigma_{\pm}$ polarization states) can be generated via a linear birefringence of the cavity material induced e.g. by a mechanical stress~\cite{stress} or by a sub-wavelength grating imprinted on the cavity~\cite{gratings}.

\begin{figure}[htbp]
\includegraphics[width=0.8\columnwidth]{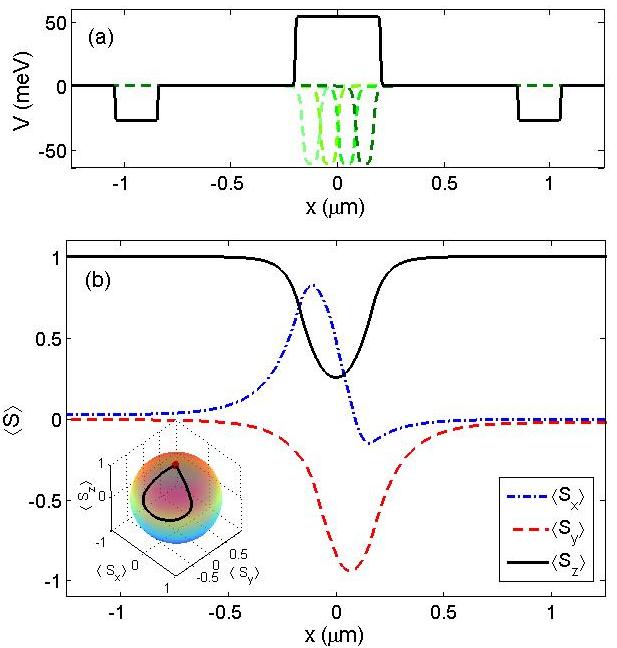}
\caption{(Color online)
Laterally patterned planar microcavity scheme to generate the artificial magnetic field.
(a) Spatial dependence of the scalar potential $V_{sc}(x)$ (solid line) and the amplitude of the $xy$ component of the vector field (dashed lines); different dashed lines correspond to the components of amplitude $V_s(x-x_j)$ centered at $x_j$ having different angles $\zeta_j$ with $\zeta_{\rm max} = 2\pi/3$. Effective photon mass $m$ was taken to be $5\times10^{-5}$ times the electron mass. The $z$ component of the vector field is $V_z(x) = -1.21$ meV. (b) Spatial dependence of the ground state expectation value of different spin components $\langle S_i \rangle=\frac{1}{2}\langle \hat{\sigma}_i \rangle$. Inset shows the corresponding loop on the Poincar\'e sphere. \label{fig:planar}}
\end{figure}

Specifically, we consider the one-dimensional configuration sketched in Fig. \ref{fig:planar}(a) that gives rise to a Hamiltonian of the form
\begin{equation}
H= 
\frac{p_x^2}{2m}+V_{sc}(x)
+ V_z(x)\,\hat{\sigma}_z
+\sum_j V_s(x-x_j)
R_{\zeta_j}^{-1}\hat{\sigma}_x R_{\zeta_j},
\label{eq:schroedinger}
\end{equation}
where $m$ is the effective photon mass along the two-dimensional cavity plane, $\hat{\sigma}$'s are the Pauli matrices in the two-dimensional spin space spanned by the $\sigma_\pm$ polarization states. $R_\zeta=\exp(-i\hat{\sigma}_z\zeta/2)$ are the rotation operators around the $z$ axis.

The scalar potential $V_{sc}(x)$ [solid line in Fig. \ref{fig:planar}(a)] stems from the lateral patterning of the cavity layer thickness and confines the photons in two square wells; the additional barrier in the center serves to cancel the localizing effect of the vector field in the $xy$ plane. The component $V_z(x)$ of the vector field coupling to the $z$ component of the effective spin is assumed to be constant in space.

The component of the vector field along the $xy$ plane is localized in between the two wells and is modeled as a superposition of several ($j=1,2,\ldots,j_{\rm max}$) localized potentials of amplitude $V_s(x-x_j)$ centered around neighboring positions $x_j$ and oriented in different directions making angles $\zeta_j$ with the $x$ direction gradually varying from $0$ to a maximum value $\zeta_{\rm max}$.

The smooth variation of the angles $\zeta_j$ is intended to ensure adiabaticity: For large enough amplitudes of the vector field, the two spin states are energetically split. As a result, the photon polarization is able to adiabatically follow the local ground state determined by the direction of the local field and traces a closed loop on the Poincar\'e sphere. On general Berry phase arguments~\cite{Berry}, we can anticipate that tunneling between the wells will involve a geometric phase which is equal to half the solid angle $\Omega$ subtended by the closed loop.

This expectation has been verified by a numerical calculation of the ground state of the Hamiltonian (\ref{eq:schroedinger}) by means of an imaginary-time evolution. The ground state is localized within the potential wells determined by the scalar potential and the local expectation value of the effective spin operator indeed follows a closed loop on the Poincar\'e sphere as shown in the inset of Fig. \ref{fig:planar}(b).
As expected, the relative phase of the ground state wavefunction in the two wells is found to be very close to the value $\phi=\Omega/2$ predicted by the adiabatic model.

A full two-dimensional lattice of wells can be obtained by repeating this building block along both directions. With a suitable tuning of the hopping phase between pairs of neighboring wells, the photon turns out to experience a non-trivial artificial gauge potential $\mathbf{A}$.

\section{Observables}
\label{sec:observables}

After having discussed possible methods of creating an effective magnetic field for photons in a lattice, we will now turn to its observable consequences. For the sake of simplicity, we shall concentrate on the case of a uniform magnetic field and vanishing photon-photon interactions.
In our calculations we consider a finite-size, two-dimensional square lattice within the tight-binding limit and we include the pumping and loss terms describing the coupling of the cavity system with the outside world in terms of a master equation in the standard form $\partial_t\rho = i[\rho,H]/\hbar+\mathcal{L}[\rho]$~\cite{Walls}.

In the case of non-interacting photons, the tight-binding Hamiltonian of the isolated system has the following single-particle form
\be
\label{Hofstadter_Hamiltonian}
H = \sum_i \hbar \omega_\circ \hat{b}^\dag_i \hat{b}_i -\hbar J\sum_{\langle i,j\rangle} \hat{b}_i^{\dag}\hat{b}_j e^{i\phi_{ij}}+ \sum_i \left[\hbar F_i(t)\, \hat{b}_i^{\dag}+\textrm{h.c.}\right],
\ee
where $\hat{b}_i^{\dag}$ ($\hat{b}_i$) is the bosonic creation (annihilation) operator for site $i$. The hopping phase $\phi_{ij}$ stems from the artificial gauge field; $\omega_\circ$ is the natural cavity frequency and $J$ is the tunneling strength between nearest neighbor sites. We assume the coherent driving term $F_i(t)$ to be monochromatic at frequency $\omega_p$ and to act on the single site $n$, $F_i(t) = \bar{F}\,\delta_{in}\,e^{-i\omega_p t}$. Photon losses at a rate $\gamma$ are included via the standard Lindblad term
\begin{equation}
\mathcal{L}[\rho]=\gamma \sum_i \big[\hat{b}_i\rho \hat{b}_i^{\dag}-(\hat{b}_i^{\dag}\hat{b}_i\rho+\rho \hat{b}_i^{\dag}\hat{b}_i)/2\big].
\end{equation}
As we are considering a non-interacting system, the state of the field is a product of coherent states on each site with an amplitude $\beta_i = \langle \hat{b}_i \rangle$ determined by the corresponding classical field evolution equations
\begin{equation}
i\dot{\beta}_i= (\omega_\circ-i\gamma/2)\,\beta_i-J \sum_{\langle j \rangle} e^{i\phi_{ij}}\,\beta_j+F_i(t),
\end{equation}
where the sum over $\langle j \rangle$ is restricted to the nearest neighbors of site $i$.

As usual in optical devices, the amplitude of the emitted light by each site is proportional to the bosonic operator $\hat{b}_i$: differently from the standard paradigm of quantum mechanics, the phase of the photonic wavefunction is then an experimentally accessible quantity, which is sensitive to the gauge potential $\mathbf{A}$ and not only to the magnetic field $\nabla\!\times\!\mathbf{A}$.

\begin{figure}[htbp]
\includegraphics[width=\columnwidth]{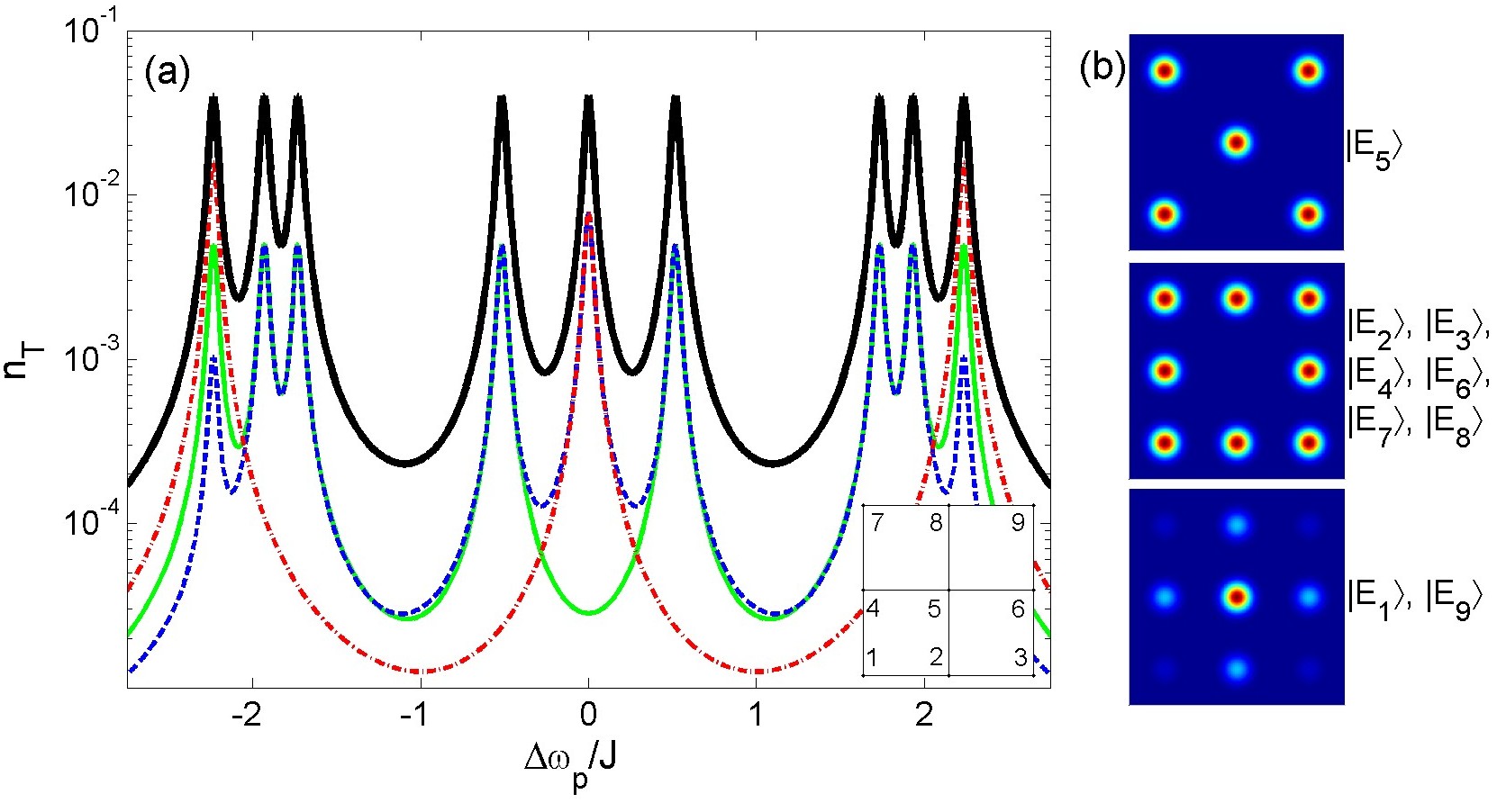}
\caption{(Color online) (a) Average total number $n_T$ of transmitted photons as a function of $\Delta\omega_p = \omega_p-\omega_\circ$. The system considered is a $3\times3$ lattice with flux quanta per plaquette $\alpha = 1/3$ and $\gamma/J = 0.05$.
Different (green, blue, red) curves correspond to a driving of amplitude $\bar{F}/J = 0.005$ localized on different (2,3,5) sites according to the enumeration given in the inset. Solid black line shows the sum over all nine sites.  (b) Real-space profile $|\psi^{(l)}(\mathbf{r})|^2$ of eigenmodes $|E_l\rangle$, where Gaussian real-space basis functions were assumed at each site for illustrative purposes. \label{fig:phi1over3}}
\end{figure}

As a specific example, we consider the case of a uniform magnetic field with the number of flux quanta per plaquette $\alpha=(2\pi)^{-1}\,\sum_{\square}\phi_{ij}$, where the sum is along a closed loop surrounding the plaquette.
The total number of photons $n_T = \sum_i \langle \hat{b}_i^\dagger \hat{b}_i \rangle$ present in the system is plotted in Fig. \ref{fig:phi1over3}(a) as a function of the pump frequency $\Delta\omega_p = \omega_p-\omega_\circ$ for a $3\times3$ lattice at $\alpha = 1/3$ with hard-wall boundary conditions, as is relevant for experiments. In the laterally patterned microcavity configuration of Fig. \ref{fig:planar}, $n_T$ is proportional to the total transmitted intensity across the system.

In Fig. \ref{fig:phi1over3}(a), the different colored curves $n_{T,(n)}$ correspond to localized driving on different sites $n$. All curves exhibit peaks at frequencies corresponding to the eigenmodes of the system. For each eigenmode, the peak strength is proportional to the weight of the eigenmode on the driven site $n$.
This physics is summarized by the explicit expression
\begin{equation}
n_{T,(n)} = \sum_i \langle \hat{b}_i^\dagger \hat{b}_i \rangle_{(n)}=|\bar{F}|^2\sum_l \frac{|\psi_n^{(l)}|^2}{(\Delta\omega_p-E_l)^2+\gamma^2/4},
\end{equation}
where $\psi_n^{(l)}$ is the component on site $n$ of the wavefunction corresponding to the eigenmode $l$ of the hopping Hamiltonian [i.e. the second term in Eq. (\ref{Hofstadter_Hamiltonian})] and $E_l$ is the corresponding eigenfrequency. Examples of the wavefunctions $\psi^{(l)}$ of different eigenmodes are shown in Fig. \ref{fig:phi1over3}(b).

Analogously, the local field on site $i$ under a localized drive on site $n$ is given by
\begin{equation}
\beta_{i,(n)}= \bar{F}\sum_l \frac{\psi_i^{(l)}\,\psi_n^{(l)\ast}}{\Delta\omega_p+i\gamma/2-E_l},
\end{equation}
which can be recognized as $\bar{F}$ times the Green's function $G(i,n,\Delta\omega_p+i\gamma/2)$ of the non-lossy system, measured at frequency $\Delta\omega_p+i\gamma/2$ and positions $i,n$. Green's function methods have been widely used in the literature to study, e.g., the effects of disorder on the Hofstadter spectrum~\cite{Green}.

\begin{figure}[htbp]
\includegraphics[width=\columnwidth]{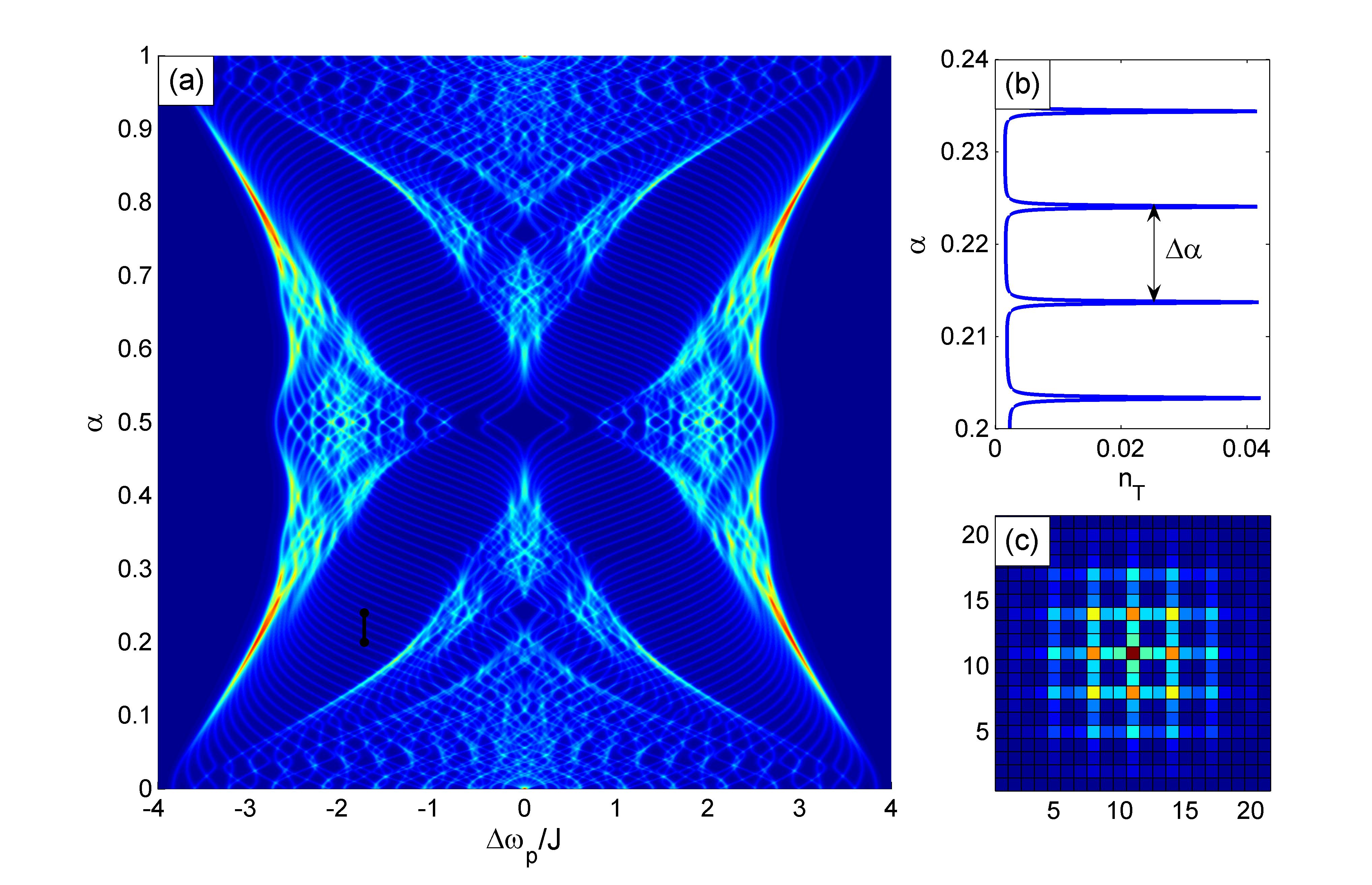}
\caption{(Color online) (a) Color plot of the total transmission $n_T$ as a function of flux quanta per plaquette $\alpha$ and pump frequency $\Delta\omega_p/J$ for a $10\times10$ lattice with $\bar{F}/J = 0.005$ and $\gamma/J = 0.05$. (b) cut of the color plot along the constant $\Delta \omega_p$ line indicated in (a) as a black line. (c) Photon occupation number $\langle \hat{b}_i^\dagger \hat{b}_i\rangle$ pattern for a pump resonant with the ground state and localized on a single site at the lattice center. $21\times21$ lattice, magnetic flux quanta per plaquette $\alpha=1/3$, $\gamma/J=0.01$.
\label{fig:3x3_10x10}}
\end{figure}

To clearly illustrate the Hofstadter butterly, it is useful to plot the sum of transmission spectra over all possible experimental realizations in which only one site is pumped at a time, as shown by the black line in Fig. \ref{fig:phi1over3}(a). It can be shown that in the limit of vanishing losses this quantity $\sum_n n_{T,(n)}$ is proportional to the density of states of the Hofstadter spectrum (modified by finite-size effects), so it is natural to expect that the main features of the spectrum will be retained in the presence of weak but finite losses.
In Fig. \ref{fig:3x3_10x10}(a), this quantity is plotted as a function of both $\Delta\omega_p$ and $\alpha$ for a $10\times10$ lattice. In addition to the clearly visible butterfly structure that closely resembles the infinite-size case, one can also recognize a series of low intensity lines appearing within the largest energy gap. Direct inspection of the eigenstates shows that these lines correspond to edge states. From Fig. \ref{fig:3x3_10x10}(b), it is apparent that the separation between neighboring lines has an almost constant value $\Delta \alpha$ approximately equal to $1/\mathcal{A}$, where $\mathcal{A}=(L-1)^2$ is the area enclosed by the outermost sites of an $L\times L$ lattice. This value of $\Delta \alpha$ corresponds to a change of magnetic flux across the whole lattice by one flux quantum and can be interpreted by the St\v{r}eda formula for the quantized Hall conductance~\cite{Streda}.

Another interesting feature of the Hofstadter physics is the spatially periodic structure of the ground state wavefunction for rational values of $\alpha$~\cite{Bhat}. In our photonic system, this can be experimentally studied by tuning the pump frequency on resonance with the lowest frequency peak and collecting transmitted light from each site in a spatially-selective way. As an example, we show in Fig. \ref{fig:3x3_10x10}(c) that for $\alpha= 1/3$ the pattern exhibits a simple periodicity of $3$ sites.

\section{Conclusions}

In this work, we have proposed two configurations where the photon experiences an artificial gauge potential in a solid-state photonic device. While the first scheme appears suitable for a first experimental demonstration of artificial gauge fields for non-interacting photons, the second one is promising in view of combining the gauge field with strong photon-photon interactions.
We have pointed out observable consequences of the artificial gauge field in the experimentally accessible optical spectra of the device and have identified clear signatures of the Hofstadter physics at strong magnetic fields. Future theoretical work will aim at extending this study to the interacting regime where novel non-equilibrium features of strongly correlated quantum Hall fluids of light are expected to appear.

\section{Acknowledgments}

We are grateful to M. Ghulinyan, P. Bettotti, M. Richard, and M. Hafezi for continuous discussions and acknowledge financial support from ERC through the QGBE grant.

\appendix
\section{Structure of the photonic modes within each cavity}

\begin{figure}[htbp]
\includegraphics[scale=0.7]{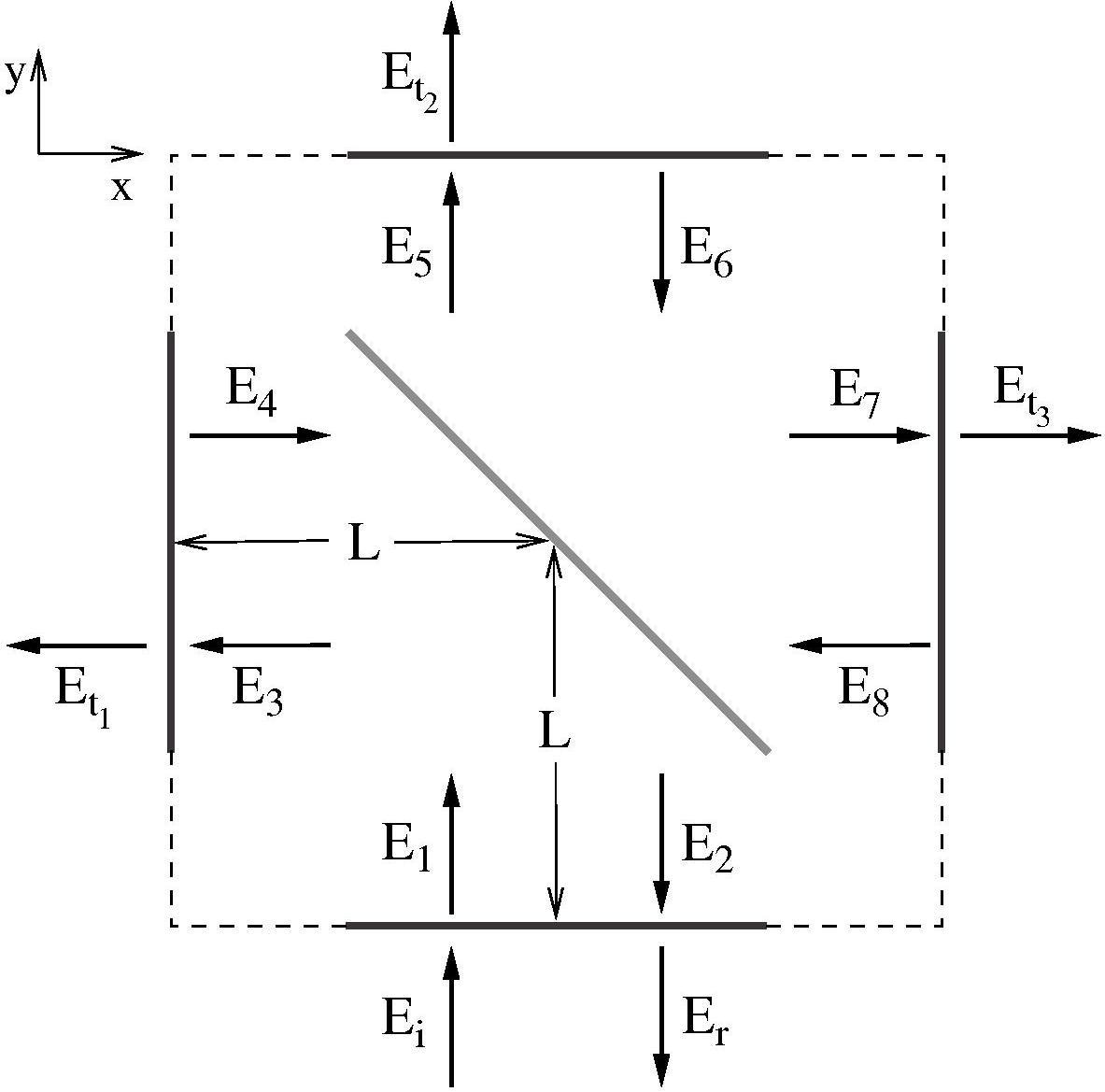}
\caption{Schematic arrangement of mirrors forming the cavity. Central mirror is located at 45 degrees with respect to $x$ and $y$ axes, its center being at a distance L from the external mirrors.
\label{mirrors}}
\end{figure}

In this Appendix, we provide more details on the structure of the photonic eigenmodes within each cavity of the two-dimensional array considered in Sec.~\ref{sec:first}. In particular, we analyze the mechanism of the coupling between light beams propagating along the $x$ and $y$ directions. The element that couples the two directions is a partially reflecting mirror inserted inside the cavity making an angle of 45 degrees with the $x$ and $y$ axes as shown in Fig. \ref{mirrors}.

We expand the electric field along each of the four arms inside the cavity as two counter-propagating plane-waves and use the appropriate scattering matrices to match the incoming and outgoing field amplitudes at the central and external mirrors. This matching yields the following matrix equations

\bea
\left(
  \begin{array}{cc}
    t & r \\
    r & t \\
  \end{array}
\right) \left(
         \begin{array}{c}
           E_i \\
           E_2 \\
         \end{array}
       \right) = \left(
         \begin{array}{c}
           E_1 \\
           E_r \\
         \end{array}
       \right)\label{firstmatrix},
\eea

\bea
\left(
  \begin{array}{cc}
    t & r \\
    r & t \\
  \end{array}
\right) \left(
         \begin{array}{c}
           0 \\
           E_3 \\
         \end{array}
       \right) = \left(
         \begin{array}{c}
           E_4 \\
           E_{t_1} \\
         \end{array}
       \right),
\eea
\bea
\left(
  \begin{array}{cc}
    t & r \\
    r & t \\
  \end{array}
\right) \left(
         \begin{array}{c}
           0 \\
           E_5 \\
         \end{array}
       \right) = \left(
         \begin{array}{c}
           E_6 \\
           E_{t_2} \\
         \end{array}
       \right),
\eea
\bea
\left(
  \begin{array}{cc}
    t & r \\
    r & t \\
  \end{array}
\right) \left(
         \begin{array}{c}
           0 \\
           E_7 \\
         \end{array}
       \right) = \left(
         \begin{array}{c}
           E_8 \\
           E_{t_3} \\
         \end{array}
       \right),
\eea
\bea
\left(
  \begin{array}{cc}
    t_\circ & r_\circ \\
    r_\circ & t_\circ \\
  \end{array}
\right) \left(
         \begin{array}{c}
           E_1 e^{i\varphi} \\
           E_8 e^{i\varphi} \\
         \end{array}
       \right) = \left(
         \begin{array}{c}
           E_5 e^{-i\varphi} \\
           E_3 e^{-i\varphi} \\
         \end{array}
       \right),
\eea
\bea
\left(
  \begin{array}{cc}
    t_\circ & r_\circ \\
    r_\circ & t_\circ \\
  \end{array}
\right) \left(
         \begin{array}{c}
           E_4 e^{i\varphi} \\
           E_6 e^{i\varphi} \\
         \end{array}
       \right) = \left(
         \begin{array}{c}
           E_7 e^{-i\varphi} \\
           E_2 e^{-i\varphi} \\
         \end{array}
       \right)\label{lastmatrix},
\eea
where $t$ ($r$) is the transmissivity (reflectivity) of the external mirrors defined as the ratio of the transmitted (reflected) amplitude to the incident amplitude, $t_\circ$ ($r_\circ$) is the transmissivity (reflectivity) of the central mirror embedded in the cavity, and $\varphi = \omega Ln/c$ is the phase that a wave gets after it travels a distance $L$ in the cavity, $n$ being the refractive index of the cavity material. The field amplitudes are defined in the immediate vicinity of the external mirrors. Here, for simplicity, we assumed that the media inside and outside of the cavity are the same and further took all mirrors to be symmetric. This condition, together with the unitarity of scattering matrices (dictated by flux conservation) is satisfied by choosing $r$ and $r_\circ$ to be purely imaginary and $t$, $t_\circ$ to be real.

So, the mirrors can now be characterized by only one parameter, as transmissivities and reflectivities are related through $r = i\sqrt{1-t^2}$ and $r_\circ = i\sqrt{1-t_\circ^2}$. The $i$ factor in the reflectivities means that upon reflection from the mirrors the fields acquire a phase factor of $e^{i\pi/2}$. The six matrix equations [Eqs. (\ref{firstmatrix}-\ref{lastmatrix})] written above provide us with twelve linear equations for the twelve unknown fields all scaled by the incident field $E_i$. In what follows we will adopt the convention that $E_i=1$.

Let us first suppose that the central mirror is perfectly transmitting, i.e. $t_\circ = 1, r_\circ = 0$. It is obvious in this case that no field will develop along the $x$ direction (i.e. $E_{3,4,7,8}=0$) as the incident field is propagating along $y$. Solving the equations one can find the transmitted field intensity to be $ |E_{t_2}|^2 = |t|^4/(1+|r|^4+2|r|^2\cos{4\varphi})$. In order for this quantity to be maximum so that we could say a cavity mode develops, the condition $\cos{4\varphi}=-1$ should be satisfied yielding \be\omega = \frac{\pi(2N+1)c}{4nL},\label{omega}\ee where $N$ is an integer. This relation can also be understood simply in terms of the round-trip condition in a cavity. For constructive interference to occur, the phase accumulated by the electric field in a round-trip [in this case $2(2L\omega n/c)+\pi$, $\pi$ being the extra phase due to reflection from the two external mirrors] must be an integer multiple of $2\pi$. When $|E_{t_2}|^2$ is plotted as a function of $\omega$, it will then exhibit peaks with magnitude one at $\omega$ values given by Eq. (\ref{omega}). The line-width of these peaks is determined by the reflectivity of the external mirrors: the higher the reflectivity the narrower the peaks.

\begin{figure}[htbp]
\includegraphics[width=\columnwidth]{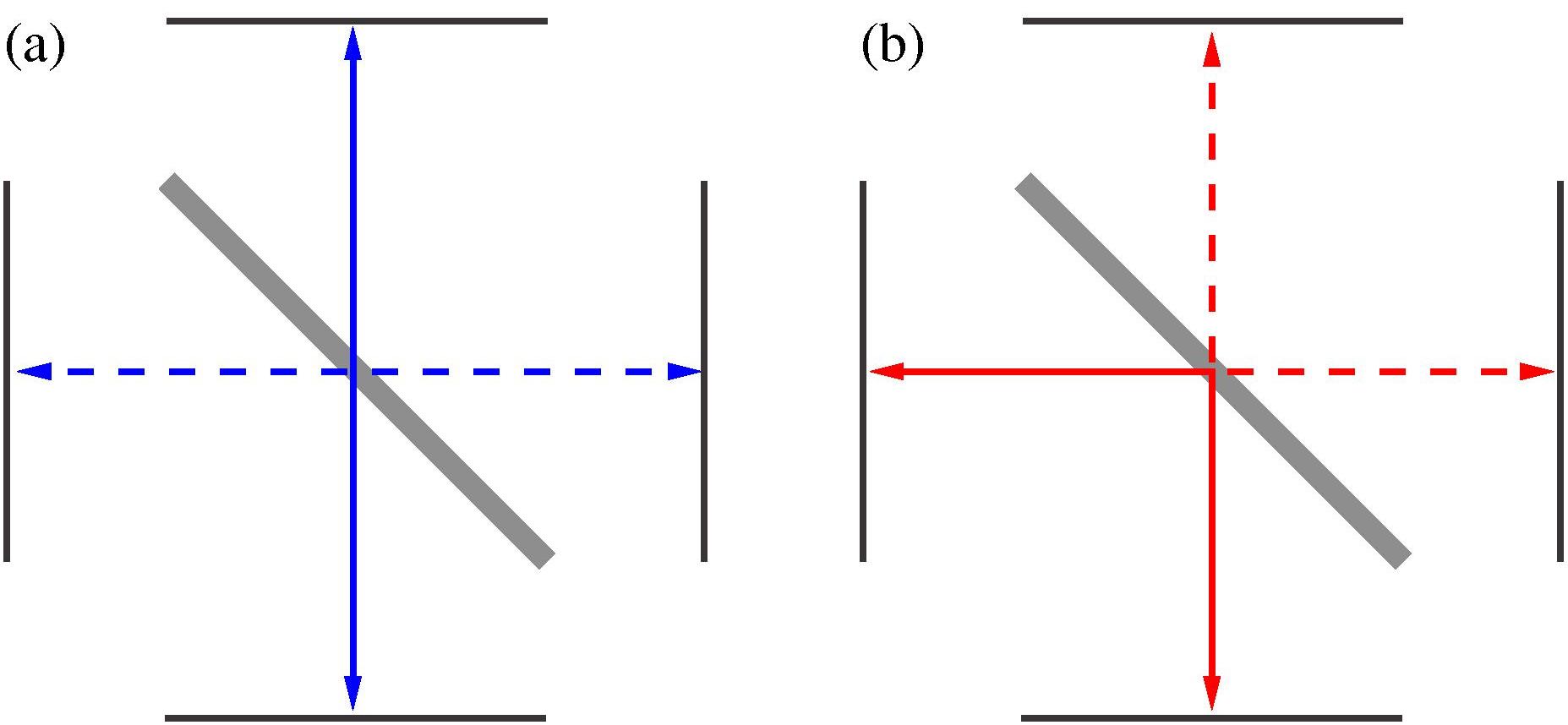}
\caption{(Color online) Solid and dashed double-headed arrows symbolize two different modes that are weakly coupled by the central mirror. (a) $|r_\circ|\ll1$, (b) $t_\circ\ll1$.
\label{modes}}
\end{figure}

What we have described up to this point is just a standard Fabry-Perot cavity. If we now increase the reflectivity of the central mirror slightly, i.e. for $|r_\circ|\ll 1$, an electric field propagating along the $x$ direction will also build up. To understand the physics of this system, one can think of the $x$ and $y$ propagating modes being weakly coupled by the central mirror. Then the eigenmodes will be symmetric and antisymmetric superpositions of these two modes, one along $x$ and one along $y$ as depicted in Fig. \ref{modes}(a). The frequency separation between symmetric and antisymmetric combinations is proportional to the amplitude of the small enough reflectivity $|r_\circ|$, which can now be thought as the coupling coefficient.

\begin{figure}[htbp]
\includegraphics[width=\columnwidth]{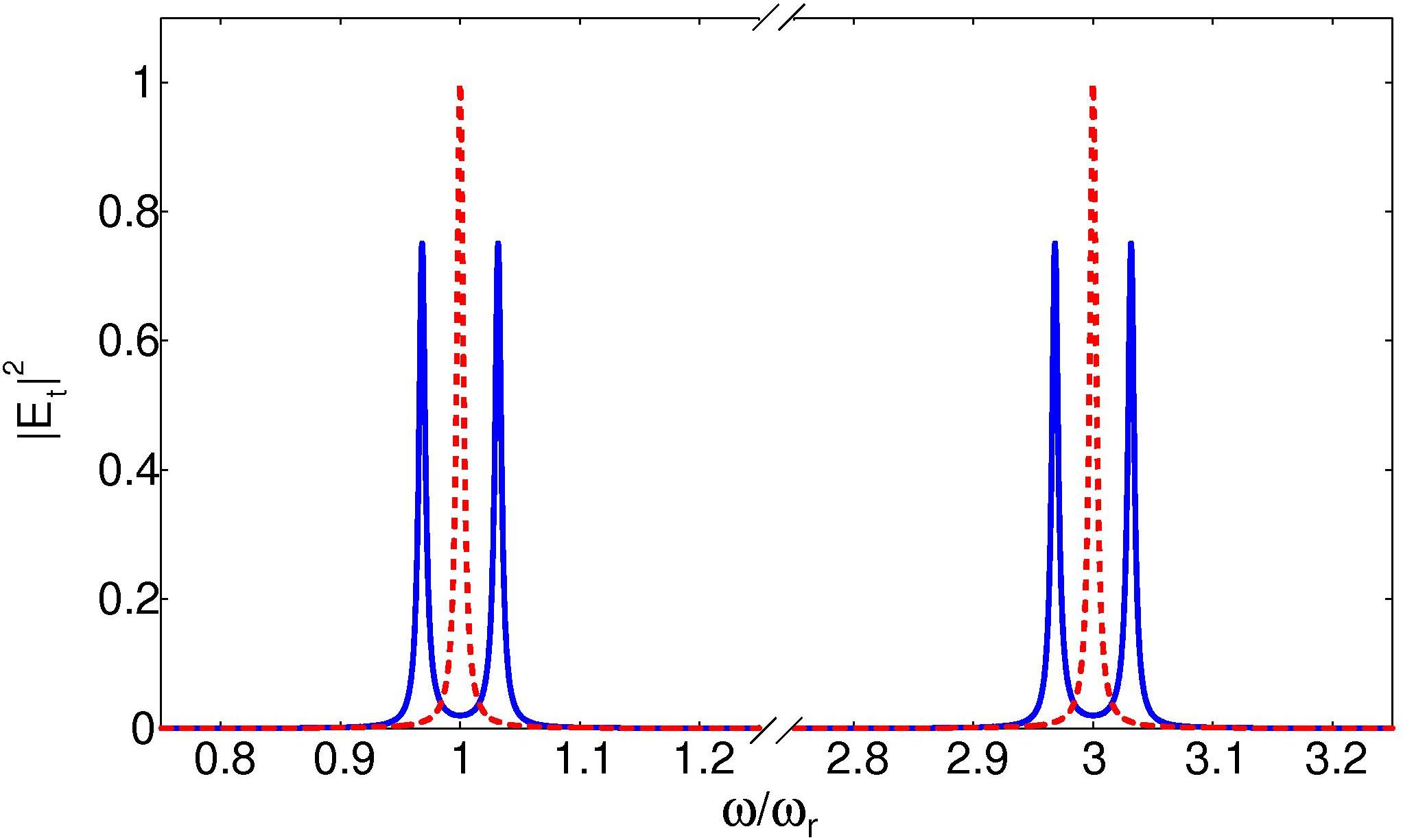}
\caption{(Color online) Total transmission $|E_t|^2=|E_{t_1}|^2+|E_{t_2}|^2+|E_{t_3}|^2$ as a function of $\omega/\omega_{\rm r}$ with $t = 0.1$ and $L = \pi c/4n\omega_{\rm r}$. Red dashed line shows the case for $|r_\circ| = 0$ and blue solid line is for $|r_\circ| = 0.05$.
\label{E_t}}
\end{figure}

\begin{figure}[htbp]
\includegraphics[width=\columnwidth]{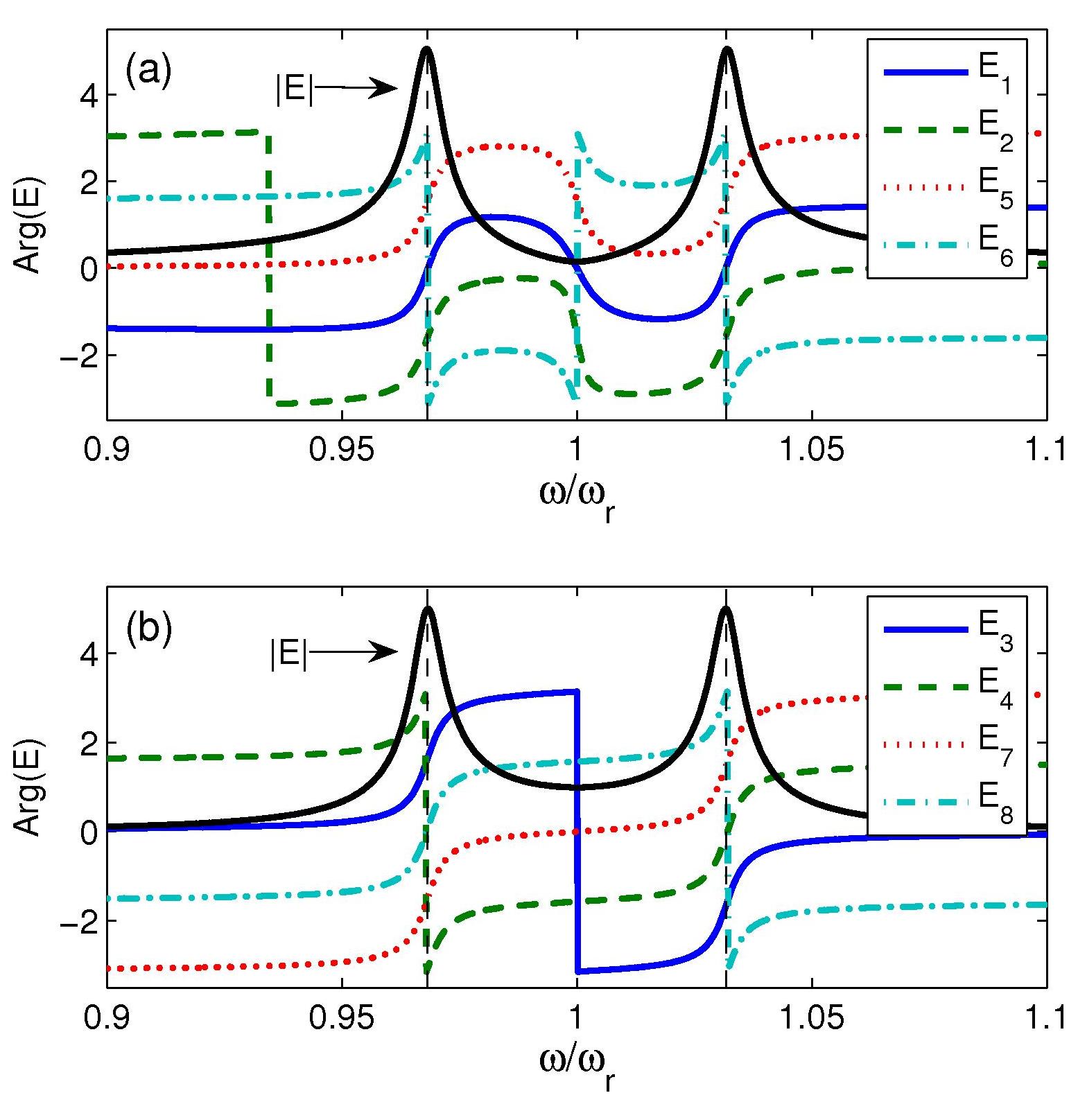}
\caption{(Color online) Magnitude and phase of the complex valued electric fields inside the cavity as a function of $\omega/\omega_{\rm r}$ for $t = 0.1$, $|r_\circ| = 0.05$, and $L = \pi c/4n\omega_{\rm r}$. Phases are defined in the range $[-\pi,\pi)$. $y$ axis can also be used to read off $|E|$. Vertical dashed lines correspond to peak positions in the total transmission of Fig. \ref{E_t}. (a) Arguments of $E_1$, $E_2$, $E_5$, and $E_6$ are shown. Only $|E_1|$ is depicted as other magnitudes behave very similarly. (b) Arguments of $E_3$, $E_4$, $E_7$, and $E_8$ along with $|E_3|$ are shown.
\label{complexE}}
\end{figure}

The preceding ideas were exemplified by a numerical calculation for which the total transmitted intensity $|E_{t_1}|^2+|E_{t_2}|^2+|E_{t_3}|^2$ as a function of $\omega$ is plotted in Fig. \ref{E_t} with $L = \pi c/4n\omega_{\rm r}$, where $\omega_{\rm r}$ is a convenient reference frequency. If $r_\circ$ were identically zero, transmission would show peaks at frequencies which are odd integer multiples of $\omega_{\rm r}$ [c.f. Eq. (\ref{omega})]. For a small reflectivity, however, one can notice that there occur two peaks around each of these frequencies. To verify that these two peaks correspond to the symmetric and antisymmetric superpositions of the two modes of Fig. \ref{modes}(a), we inspected the region around $\omega/\omega_{\rm r}=1$ more closely and plotted the magnitude and phase of the electric fields inside the cavity in Fig. \ref{complexE}.

One observation is that the peak positions and magnitudes at these peaks are nearly the same for all electric field amplitudes. Nevertheless, one can identify two groups of fields in respect of their behavior as $\omega$ varies. The first group consists of $E_1$, $E_2$, $E_5$, and $E_6$ contributing to the $y$ mode of Fig. \ref{modes}, and the second group is composed of $E_3$, $E_4$, $E_7$, and $E_8$ building up the $x$ mode. One can notice from Fig. \ref{complexE} that while the phases of a field in the first group are nearly the same at both peaks, the phase of a field in the second group at the higher frequency peak is shifted almost by $\pi$ with respect to its value at the lower frequency peak. This shows that the higher (lower) frequency peak corresponds to an antisymmetric (symmetric) combination of the $x$, $y$ modes.

Similar arguments can be advanced for the case $t_\circ\ll1$, i.e. when the central mirror is weakly transmitting. We can again describe the system in a two-mode approximation and regard $t_\circ$ as the coupling coefficient. As shown in Fig. \ref{modes}(b), in this case the modes are confined in the lower and upper triangular regions of the cavity separated by the central mirror. This time peaks will appear around $\omega = \pi Nc/2nL$ as opposed to the value given by Eq. (\ref{omega}). This change is due to the phase acquired upon reflection from the central mirror.

Two final remarks are in order.\\
(i) As our scheme depends on the preservation of the type of circular polarization (defined in terms of angular momentum) inside the cavities, the central mirror should be chosen in a way that it reflects the two linear TE and TM polarizations almost equally and without introducing any appreciable relative phase: this condition may be non-trivial to obtain for light at an oblique incidence. Still, any residual mixing of the $\sigma_\pm$ polarized eigenmodes of the cavity can be suppressed by lifting their degeneracy, e.g. by introducing a slab of circularly birefringent medium in the cavity.\\
(ii) Although we have restricted ourselves to a single cavity case in this Appendix, the same formalism can be generalized to several coupled cavities. As shown in~\cite{Hafezi}, this leads to an alternative way of characterizing the optical response of a many-cavity system which does not rely on a tight-binding Hamiltonian such as (\ref{Hofstadter_Hamiltonian}).


\begin{thebibliography}{99}

\bibitem{FQH_review}
S. das Sarma and A. Pinczuk, eds., {\em Perspectives in Quantum Hall Effects} (Wiley, New York, 1997).
\bibitem{Hofstadter} D. R. Hofstadter, Phys. Rev. B {\bf 14}, 2239 (1976).
\bibitem{Albrecht} The splitting of Landau levels was observed in a two dimensional electron gas under a weak superlattice potential in C. Albrecht \textit{et al.}, Phys. Rev. Lett. \textbf{86}, 147 (2001).
\bibitem{Berry} A. Shapere and F. Wilczek, eds., {\em Geometric Phases In Physics} (World Scientific, Singapore, 1989).
\bibitem{olshanii} R. Dum and M. Olshanii, Phys. Rev. Lett. {\bf 76}, 1788 (1996).
\bibitem{Dalibard} J. Dalibard \textit{et al.}, arXiv:1008.5378 (2010).
\bibitem{spielman} Y.-J. Lin {\em et al.}, Nature {\bf 462}, 628 (2009).
\bibitem{FQH} N. R. Cooper {\em et al.}, Phys. Rev. Lett. {\bf 87}, 120405 (2001);
C.-C. Chang {\em et al.}, Phys. Rev. A {\bf 72}, 013611 (2005);
M. Hafezi \textit{et al.}, \textit{ibid.} {\bf 76}, 023613 (2007);
R. N. Palmer \textit{et al.}, \textit{ibid.} {\bf 78}, 013609 (2008).
\bibitem{EPN} I. Carusotto and C. Ciuti, Europhysics News, Vol. 41, No. 5, 23 (2010).
\bibitem{PhotBEC} J. Kasprzak {\em et al.}, Nature {\bf 443}, 409 (2006).
\bibitem{Pol_Superfl_Exp} A. Amo {\em et al.}, Nature {\bf 457}, 291 (2009); A. Amo {\em et al.}, Nature Phys. {\bf 5}, 805 (2009).
\bibitem{Pol_Vort_Exp} A. Amo {\em et al.}, Science {\bf 332}, 1167 (2011); G. Nardin {\em et al.},     Nature Physics {\bf 7}, 635 (2011); D. Sanvitto {\em et al.}, arXiv:1103.4885 (2011).
\bibitem{atoms} J. Cho \textit{et al.}, Phys. Rev. Lett. {\bf 101}, 246809 (2008);
\bibitem{PC} F. D. M. Haldane and S. Raghu, Phys. Rev. Lett. {\bf 100}, 013904 (2008);
Z. Wang {\em et al.}, Nature {\bf 461}, 772 (2009).
\bibitem{Girvin} J. Koch \textit{et al.}, Phys. Rev. A {\bf 82}, 043811 (2010); A. Nunnenkamp \textit{et al.}, arXiv:1105.1817 (2011).
\bibitem{Hafezi} During the review process of the present paper, we became aware of related works in M. Hafezi \textit{et al.}, arXiv:1102.3256 (2011) and J. Keeling, arXiv:1106.0682 (2011).
\bibitem{TG} D. E. Chang {\em et al.}, Nature Physics {\bf 4}, 884 (2008);
M. H. Hartmann {\em et al.}, Laser \& Photon. Rev. {\bf 2}, 527 (2008); I. Carusotto {\em et al.}, Phys. Rev. Lett. 103, 033601 (2009).
\bibitem{Berry_optics} A. Tomita and R. Y. Chiao, Phys. Rev. Lett. {\bf 57}, 937 (1986);
R. Y. Chiao {\em et al.}, {\em ibid.} {\bf 60}, 1214 (1988).
\bibitem{pancha} S. Pancharatnam, Proc. Ind. Acad. Sci. A {\bf 44}, 247 (1956);
M. V. Berry, Journal of Modern Optics {\bf 34}, 1401 (1987);
T. H. Chyba {\em et al.}, Optics Letters {\bf 13}, 562 (1988);
Z. Bomzon {\em et al.}, Optics Letters {\bf 27}, 1141 (2002).
\bibitem{transm_mat}  H. Benisty and J. M. G\'erard, eds., {\em Confined Photon Systems} (Springer-Verlag, Berlin, Heidelberg, 1999).
\bibitem{footnote0} Throughout the paper, $\sigma_\pm$ polarizations are defined in terms of angular momentum (which is conserved at normal incidence on a mirror) and not of helicity (which is defined as the projection of the angular momentum along the propagation wavevector, and therefore changes sign upon reflection).
\bibitem{footnote1} With {\em circularly birefringent} material, we mean an optical medium with different refractive indices for light of $\sigma_\pm$ angular momentum with respect to a given axis, independently of the propagation direction: this property underlies the Faraday rotation effect under a strong magnetic field. With {\em linearly birefringent} material, we mean an optical medium with different refractive indices for light linearly polarized along orthogonal directions. With {\em optically active} material, we mean a chiral optical medium, e.g. a sugar solution, in which the refractive index depends on the helicity.
\bibitem{patterned_cav} D. Lu {\em et al.}, Appl. Phys. Lett. {\bf 87}, 163105 (2005);
O. El Da\"if {\em et al.}, {\em ibid.} {\bf 88}, 061105 (2006);
D. Bajoni {\em et al.}, Phys. Rev. Lett. {\bf 100}, 047401 (2008).
\bibitem{blockade} A. Verger {\em et al.}, Phys. Rev. B {\bf 73}, 193306 (2006); I. Carusotto {\em et al.}, Europhys. Lett. {\bf 90}, 37001 (2010).
\bibitem{local_magnetic} O. Gaiffe {\em et al.}, Eur. Phys. J. Appl. Phys. {\bf 47}, 11201 (2009).
\bibitem{stress} A. K. Jansen van Doorn {\em et al.}, Appl. Phys. Lett. {\bf 69}, 3635 (1996).
\bibitem{gratings} D. C. Flanders, Appl. Phys. Lett. {\bf 42}, 492 (1983).
\bibitem{Walls} D. F. Walls and G. J. Milburn, {\em Quantum Optics}, 2nd edition, (Springer-Verlag, Berlin, Heidelberg, 2008).
\bibitem{Green} R. R. Gerhardts {\em et. al}, Phys. Rev. B {\bf 43}, 5192 (1991); D. Pfannkuche and R. R. Gerhardts, Phys. Rev. B {\bf 46}, 12606 (1992); U. Wulf and A. H. MacDonald, Phys. Rev. B {\bf 47}, 6566 (1993); Zhou {\em et. al}, Phys. Rev. B {\bf 71}, 125310 (2005);
\bibitem{Streda} P. St\v{r}eda, J. Phys. C {\bf 15} L717 (1982); D. J. Thouless {\it et al.}, Phys. Rev. Lett. {\bf 49}, 405 (1982); R. Rammal {\it et al.}, Phys. Rev. B {\bf 27}, 5142 (1983); K. Czajka {\it et al.}, Phys. Rev. B {\bf 74}, 125116 (2006).
\bibitem{Bhat} R. Bhat \textit{et al.}, Phys. Rev. A {\bf 76} 043601 (2007).
\end{thebibliography}
\end{document}